\documentclass[useAMS,usenatbib,referee]{mn2e}

\usepackage{graphicx}
\usepackage{amsmath}
\usepackage{bm}
\usepackage{epsfig}
\usepackage{amssymb}
\usepackage{natbib}
\usepackage{threeparttable} 
\usepackage{subfigure}
\usepackage{setspace}
\usepackage{multirow}    
\usepackage{appendix}
\newcommand\apj{ApJ}
\newcommand\apjl{ApJ}
\newcommand\apjs{ApJS}
\newcommand\apss{Ap\&SS}
\newcommand\aap{A\&A}
\newcommand\mnras{MNRAS}
\def\simmore{\mathbin{\lower 3pt\hbox
     {$\rlap{\raise 5pt\hbox{$\char'076$}}\mathchar"7218$}}}   

\title[Mhz QPOs and burst convexity in 4U 1636--53]{Millihertz quasi-periodic oscillations in 4U 1636$-$53 associated with bursts with positive convexity only}
\author[Ming Lyu et al.]
{Ming Lyu$^1$\thanks{E-mail: m.lyu@astro.rug.nl}, Mariano M\'endez$^1$, Diego Altamirano$^2$ and Guobao Zhang$^3$ \\
$^1$Kapteyn Astronomical Institute, University of Groningen, PO BOX 800, NL-9700 AV Groningen, the Netherlands\\
$^2$School of Physics and Astronomy, University of Southampton, Southampton, SO17 1BJ, UK \\
$^3$New York University Abu Dhabi, P.O. Box 129188, Abu Dhabi, United Arab Emirates
}

\begin{document}

\date{Accepted XXXX. Received XXXX; in original form XXXX}

\maketitle

\label{firstpage}

\begin{abstract}

We investigated the convexity of all type I X-ray bursts with millihertz quasi-periodic oscillations (mHz QPOs) in 4U 1636--53 using archival observations with the Rossi X-ray Timing Explorer. We found that, at a $3.5 \sigma$ confidence level, in all 39 cases in which the mHz QPOs disappeared at the time of an X-ray burst, the convexity of the burst is positive. The convexity measures the shape of the rising part of the burst light curve and, according to recent models, it is related to the ignition site of bursts on the neutron-star surface. This finding suggests that in 4U 1636$-$53 these 39 bursts and the marginally-stable nuclear burning process responsible for the mHz QPOs take place at the neutron-star equator. This scenario could explain the inconsistency between the high accretion rate required for triggering mHz QPOs in theoretical models and the relatively low accretion rate derived from observations.

\end{abstract}


\section{Introduction}  

Nearly half of the accreting neutron stars in low-mass X-ray binaries show Type I X-ray bursts \citep[e.g.,][]{intzand04,liu07,galloway08}. These bursts are due to unstable thermonuclear burning of accumulated hydrogen and helium on the surface of the neutron star \citep[e.g.,][]{fujimoto81}. In the last decade another observational phenomenon connected to nuclear burning on the neutron-star surface has been discovered. \citet{revni01} reported the first detection of quasi-periodic oscillations (QPOs) in the millihertz (mHz) range in three neutron-star low-mass X-ray binaries (NS LMXBs): 4U 1608--52, 4U 1636--53, and Aql X-1. Besides the low frequency range between 7 and 9 mHz, the mHz QPOs show some unique properties compared to other types of QPOs in NS LMXBs: The mHz QPOs happen only within a particular luminosity range, $L_{\rm 2-20\:keV} \simeq (5-11) \times 10^{36}$ ergs s$^{-1}$, and are stronger at low photon energies ($E$ $<$ 5 keV) \citep{revni01,diego08}.

\citet{diego08} found that the frequency of the mHz QPO in 4U 1636--53 decreased systematically with time until the QPO became undetectable at the time of a type I X-ray burst when the source was in the transition between hard and soft state usually seen in these systems. \citet{linares10} found mHz QPOs in the neutron-star transient IGR J17480--2446 in the globular cluster Terzan 5. These mHz QPOs showed some different properties with respect to the ones in other sources: The QPO frequency was relatively low, always below 4.5 mHz, and the persistent source luminosity at the time the QPOs appeared was high, $L_{\rm 2-50\:keV}$ $\sim 10^{38}$ erg s$^{-1}$. Furthermore, \citet{linares12} found a smooth evolution between X-ray bursts and mHz QPOs in IGR J17480--2446 as the luminosity of the source changed during the outburst, which has never been observed in other mHz QPO sources.

The above observational findings suggest a different origin of the mHz QPOs from other kinds of QPOs \citep[e.g.,][]{Straaten02,van05,Vanderklis06,diego08b} in NS LMXBs. \citet{revni01} speculated that a special mode of nuclear burning on the neutron-star surface may be responsible for the mHz QPOs. \citet{heger07} proposed that the mHz QPOs could be a consequence of marginally stable nuclear burning of Helium on the neutron-star surface. The model of \citet{heger07} is able to explain the characteristic time scale of $\sim$ 2 minutes of the mHz QPOs, and predicts that the QPOs should occur only in a very narrow range of X-ray luminosity. However, the accretion rate at which the mHz QPOs are predicted in the model is close to the Eddington rate, up to one order of magnitude higher than the one implied by the X-ray luminosity at which mHz QPOs were observed. To bring the models and observations into agreement, \citet{heger07} proposed that the local accretion rate in the burning layer where the QPOs happen can be higher than the global accretion rate. \citet{keek09} found that turbulent chemical mixing of the fuel, together with a higher heat flux from the crust, can explain the observed accretion rate at which mHz QPOs are seen. Furthermore, \citet{diego08} and \citet{keek09} suggested that the cooling process of the layer where the mHz QPOs happen may be responsible for the frequency drift of the QPOs before X-ray bursts. \citet{keek14} explored the influence of the fuel composition and nuclear reaction rates on the mHz QPOs, and concluded that no allowed variation in the composition and the reaction rate is able to trigger the mHz QPOs at the observed accretion rates.   

\citet{lyu15} investigated the relation between the frequency of the mHz QPOs and the temperature of the neutron-star surface in 4U 1636--53 using XMM-Newton and simultaneous RXTE observations, and they found that there was no significant correlation, which is different from theoretical predictions. Besides, \citet{lyu15} found that all seven X-ray bursts associated with mHz QPOs in this source were bright, energetic and short, indicating a potential connection between the mHz QPOs and He-rich X-ray bursts.

\citet{cooper07} found that the latitude at which type I X-ray bursts ignite on the neutron star surface depends on the accretion rate. Later, \citet{Maurer08} simulated the influence of ignition latitude, accretion rate and neutron-star rotation on the shape of the rising phase of type I X-ray bursts. They found that bursts that ignite at the equator always have positive convexity, whereas bursts that ignite at high latitude have both positive and negative convexity. The convexity measures the shape of the rising part of the burst light curve, and it is defined as the integrated area of the burst light curve above (positive convexity) or below (negative convexity) a straight line drawn from the start to the peak of the burst. Recently, \citet{mah15} further confirmed that the rising part of the light curve of bursts is more concave when ignition starts near the pole compared to when it starts near the equator.  Thus, the convexity of an X-ray burst provides information about the ignition site of unstable nuclear burning on the neutron-star surface. The fact that mHz QPOs are closely related to type I X-ray bursts opens up the possibility to study the origin and physics of marginally stable nuclear burning on the neutron-star surface, by investigating mHz QPOs and type I X-ray bursts together. In this paper we focus on the possible connection between mHz QPOs and the convexity of type I X-ray bursts to explore the site on the neutron-star surface at which the marginally stable nuclear burning ignites.

\section{Observations and data reduction}
We analysed all available data of 4U 1636--53 from the Proportional Counter Array \citep[PCA;][]{jahoda06} on board of the Rossi X-ray Timing Explorer (RXTE) using the Heasoft 6.16. An RXTE observation typically covers 1 to 5 consecutive 90-minute satellite orbits. Usually, an orbit contains between 1 and 5 ks of useful data separated by 1--4 ks data gaps; on rare occasions the visibility windows were such that RXTE continuously observed the source for up to $\sim$27 ks. This means that our datasets consist of continuous data segments of lengths between 0.3 and 27 ks. 

We used 1-s resolution event mode PCA light curves in the $\sim 2-5$ keV range \citep[where the mHz QPOs are the strongest, see][]{diego08} and searched for periodicities in each of the gap-free segments separately using Lomb-Scargle periodograms \citep{lomb76,scargle82,Numerical-Recipes}. In those cases where more than one Type I X-ray burst was detected, we searched for mHz QPOs before, after and in-between bursts. We only report those detections that are at least $3\sigma$ significant as estimated using the method outlined in \citet{Numerical-Recipes}. When undetected, It is difficult to estimate a general  and/or meaningful upper limit on the fractional rms amplitude of the mHz QPOs before an X-ray burst. The reasons could be many: data-gaps just before the burst, or the segment before the burst is too short to detect the QPO significantly, or there is a reduced number of PCUs during that observation. In the few cases without the above problems, we estimated $3\sigma$ upper limits as low as 0.4\% rms in the 2-5 keV range.

We investigated all X-ray bursts of 4U 1636--53 detected by the PCA/RXTE. For this we produced 0.25-s light curve from the Standard-1/Event
data and searched for X-ray bursts in these light curves following the
procedure described in \cite{Zhanggb11}.  In order to study the shape and time-scale
of the bursts rise, we extracted the bursts light curves from the PCA data with
0.125-s time resolution. To describe the shape of the burst rising phase
quantitatively, we used the convexity, $\mathcal{C}$, parameter in our analysis
\citep{Maurer08}. The convexity describes the curvature
of the light-curve rise, and it quantifies whether the curve is convex
($\mathcal{C} > 0$) or concave ($\mathcal{C} < 0$). We used the same method as in \citet{Maurer08} to calculate the convexity in the burst light curve of the full PCA energy band; Since different bursts have different durations and peak intensities, we normalised the light curves and time axes so that, from the start to the peak, each burst rises from 0 to 10 normalised intensity units within 0 to 10 normalised time units, and we calculated the convexity in the time interval in which the light curve of the burst rises from 1 to 9 normalised intensity units, for more details of the calculation please refer to \citet{Zhanggb16}. We also calculated the rising time of each burst, defined as the time interval at the beginning of a burst during which the flux in the light curve is between 10\% and 90\% of the flux at the peak of the burst.

\section{Results}
\begin{figure}
\center
\includegraphics[height=0.35\textwidth]{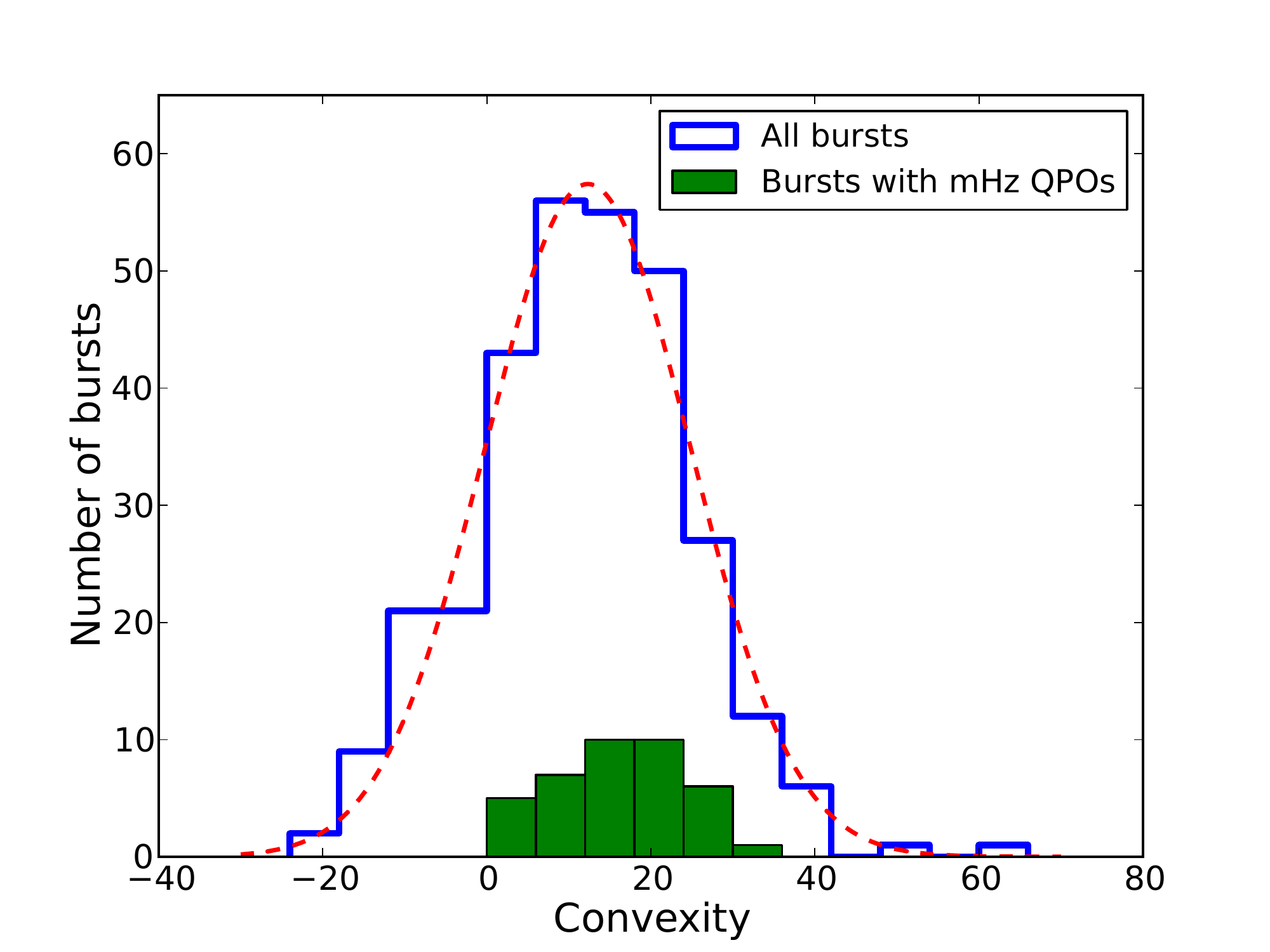}
\caption{Distribution of the convexities of all X-ray bursts (blue line) and the bursts with mHz QPOs (filled green bars) in 4U 1636--53. The red dashed line in the plot corresponds to the best-fitting Gaussian curve to the the convexity distribution of all bursts.}
\label{hist}
\end{figure}

We detected 207 cases of mHz QPOs and 371 X-ray bursts in the whole RXTE archive. We excluded, and did not analyse further, those bursts that showed at least one of the following characteristics: (i) The burst light curve was incomplete, (ii) the burst light curve had multiple peaks, or (iii) the burst was very weak and hence the light curve was very noisy. We further excluded the superburst in this source \citep{superburst,Strohm02}. We were then left with 305 burst with a complete and smooth profile. We considered that a mHz QPO and an X-ray burst are associated if, in an observation, there is a mHz QPO that ends at the same time that an X-ray burst happens. In the rest of the paper we only considered those cases in which the mHz QPOs are associated to an X-ray burst. We detected both mHz QPOs and an associated type I X-ray bursts in 39 observations; the QPOs in these observations always disappeared at the time when the associated X-ray burst appeared. In Figure \ref{hist} we show the distribution of the convexity of all type I X-ray bursts and the distribution of the convexity of those bursts that are associated with mHz QPOs in 4U 1636--53. The distribution of the convexity of all bursts is symmetric, with 252 and 53 of them having, respectively, positive and negative convexity. The distribution can be well fitted with a Gaussian function (R-square=0.976) with a mean convexity of $12.3\pm 1.2$ (95\% confidence level) and a standard deviation of $12.6\pm 1.2$.
For the 39 bursts associated with mHz QPOs, the convexities are always positive. We list the convexities of these 39 bursts in the Table \ref{convexity}. We found no case in our sample of a mHz QPO that is associated with a burst with negative convexity. In a few observations there is a second burst a few thousand seconds after the burst that is directly associated with the mHz QPO; in these cases we found that the convexity of the second burst can be either positive or negative. Furthermore, we found that the observed continuum flux ranges from 1.9 $\times$10$^{-9}$ ergs cm$^{-2}$ s$^{-1}$ to 5 $\times$10$^{-9}$ ergs cm$^{-2}$ s$^{-1}$ for bursts with mHz QPOs, and from 1.2 $\times$10$^{-9}$ ergs cm$^{-2}$ s$^{-1}$ to 8.2 $\times$10$^{-9}$ ergs cm$^{-2}$ s$^{-1}$ for bursts without mHz QPOs. The K-S test probability that the above two samples come from the same parent population is P$_{K-S}$ = 0.0083, which indicates that the distribution of the persistent flux of the two samples is marginally different.

The results shown in Figure \ref{hist} suggest that there is a relation between the presence of the mHz QPOs and the convexity of the associated burst. In order to quantify this, we calculated the probability, P$_{39}$, of selecting 39 random bursts from the distribution of the convexity of all bursts in 4U 1636--53 (see Figure \ref{hist}), and getting only bursts with positive convexity. Since the convexity can either be positive or negative, we can estimate this probability from the binomial distribution, where the probability of success (where success means $\mathcal{C} > 0$) is $P=252/305=0.826$. The probability is then P$_{39}$ =0.826$^{39}$=5.8$\times$10$^{-4}$. 

We also used the distribution in Figure \ref{hist} to simulate 10$^{6}$ sets of 39 convexities, and counted the number of trials, N$_{+}$, in which all 39 convexities were positive. We found that N$_{+} =$594, corresponding to a probability of 5.9$\times$10$^{-4}$, consistent with the calculation above.

\begin{figure}
\center
\includegraphics[height=0.4\textwidth,,angle=-90]{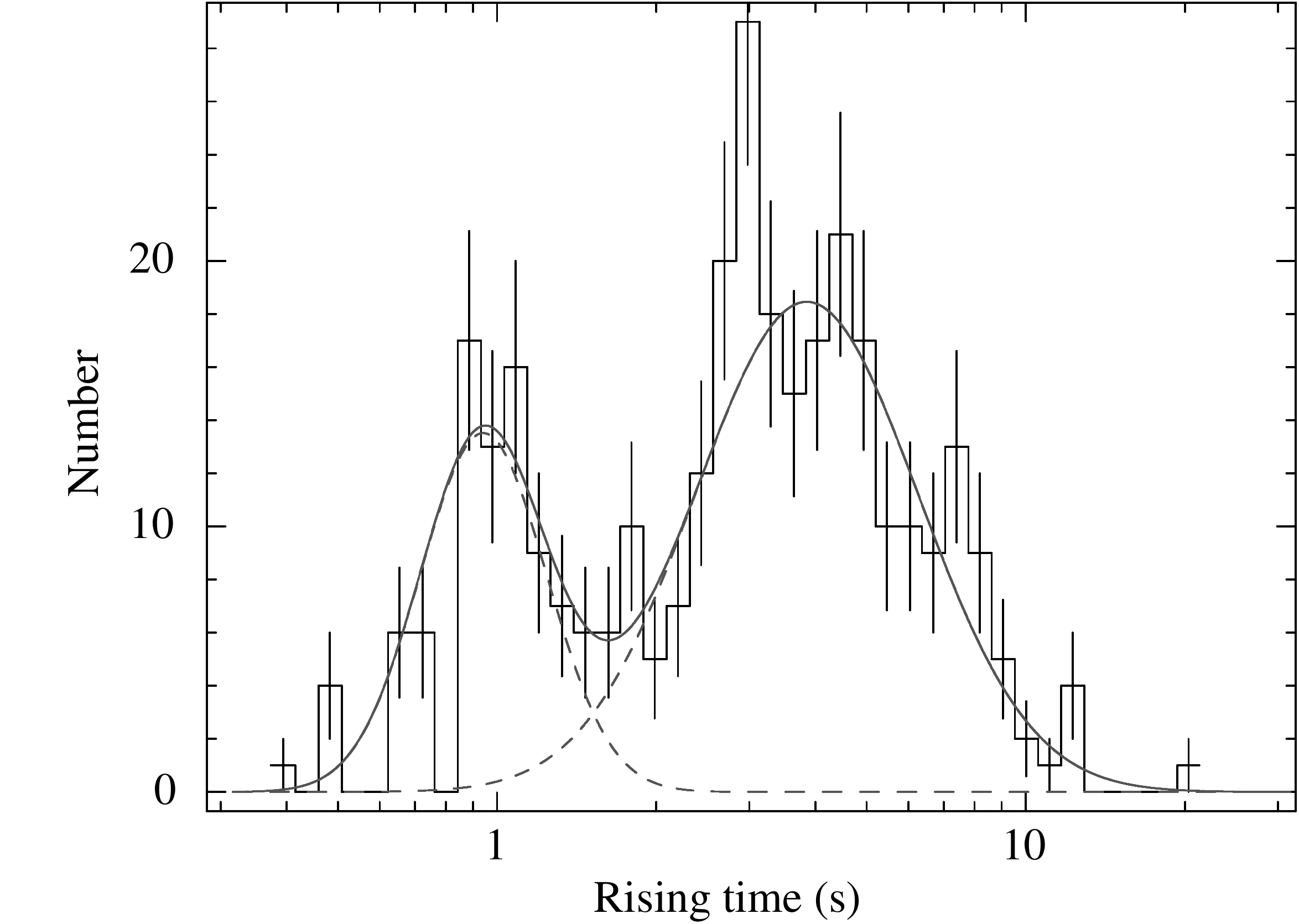}
\caption{Distribution of the rising time of X-ray bursts in 4U 1636--53. We used the dashed-lines to show the two best-fitted gaussians to the histogram, and the sum of the two Gauss components is shown as the black curve in the plot.}
\label{hist_rising}
\end{figure}

\begin{figure}
\center
\includegraphics[height=0.4\textwidth]{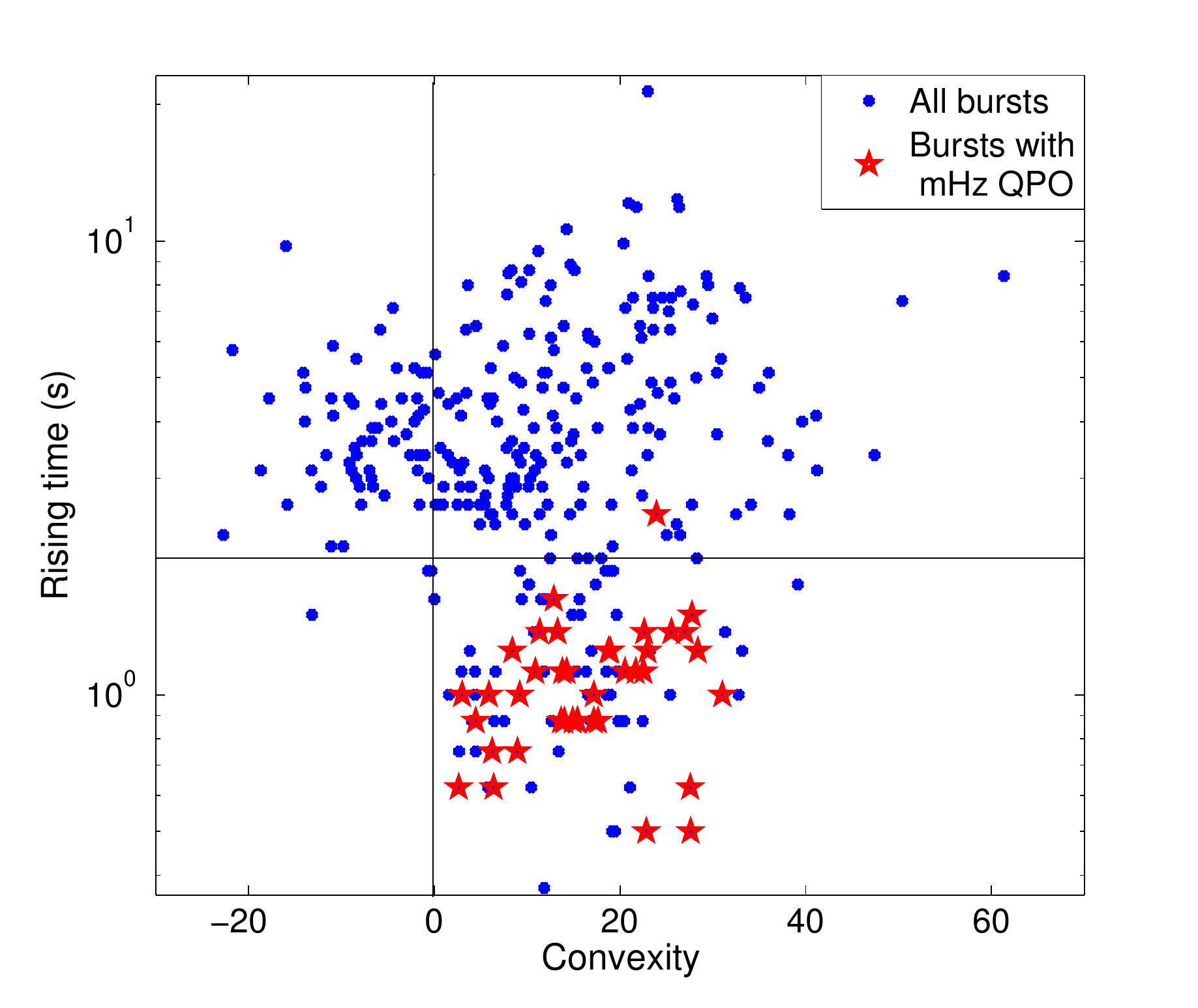}
\caption{Rising time vs. convexity of X-ray bursts (blue snow symbols) and the bursts with mHz QPOs (red stars) in 4U 1636--53. The vertical and horizontal line in the plot corresponds to a convexity equal to 0 and a rising time equal to 2 s.}
\label{rising}
\end{figure}

In Figure \ref{hist_rising} we show the distribution of the rising time of all X-ray bursts in 4U 1636--53. The rising time ranges from 0.4 s to 23 s and follows a bimodal distribution with peaks at $\sim$1 s and $\sim$3 s, respectively. In Figure \ref{rising} we show the rising time vs. convexity of all bursts (blue snow symbols) and those bursts with mHz QPOs (red stars). The vertical line in this Figure is at a convexity of zero, while the horizontal line is at a rising time equal to 2 s; the latter is approximately the value at which the distribution of rising times in Figure \ref{hist_rising} shows a local minimum. It is apparent that all bursts with mHz QPOs are located on the lower right corner of this Figure: All bursts with mHz QPO have positive convexity and, except for one case, they all have rising times shorter than 2 s. From this Figure it is also apparent that not all bursts in that part of the diagram show mHz QPOs.

\begin{table*}
\caption{List of the convexities of the 39 bursts associated with mHz QPOs in 4U 1636--53. The continuum flux is from 2 to 50 keV in unit of 10$^{-9}$ ergs cm$^{-2}$ s$^{-1}$. The convexity error here is at 1-$\sigma$ significance level.
}
\begin{tabular}{|l|c|c|c|c|c|}

\hline
ObsId   &   Star time of burst  & End time of burst   &   Convexity    & Rising time (s) & Absorbed continuum flux \\
\hline

     10088-01-08-030   &      50448.73395  & 50448.73699       &        11.3 $\pm$   1.1   	&	 1.4		   &  4.8          \\
      30053-02-02-02   &      51044.48934  & 51044.48976       &        13.3 $\pm$   1.2   	&	 1.4		   &  3.9          \\
      40028-01-02-00   &      51236.36632  & 51236.36671       &        17.2 $\pm$   1.6   	&	 1.0		   &  4.1          \\
      40028-01-04-00   &      51297.07198  & 51297.07243       &         4.5 $\pm$   1.5   	&	 0.9		   &  4.6          \\
      40028-01-08-00   &      51347.98825  & 51347.98866       &        25.5 $\pm$   1.5   	&	 1.4		   &  5.0          \\
      40031-01-01-06   &      51350.79575  & 51350.79613       &        12.9 $\pm$   1.2   	&	 1.6		   &  3.8          \\
      40028-01-15-00   &      51710.21233  & 51710.21290       &         6.4 $\pm$   2.1   	&	 0.6		   &  4.3          \\
      40028-01-19-00   &      51768.98081  & 51768.98125       &        14.2 $\pm$   1.6   	&	 1.1		   &  3.8          \\
      40028-01-20-00   &      51820.98111  & 51820.98157       &         5.9 $\pm$   1.3   	&	 1.0		   &  3.7          \\
      50030-02-05-00  &       51942.10024 &  51942.10065       &          3.0 $\pm    1.4 $	&	 1.0		   &  4.1          \\
     50030-02-09-000   &      52004.71326  & 52004.71366       &         8.4 $\pm$   1.3   	&	 1.3		   &  4.1          \\
      50030-02-10-00   &      52029.22818  & 52029.22864       &        10.9 $\pm$   1.3   	&	 1.1		   &  3.1          \\
     60032-01-02-00G   &      52075.13477  & 52075.13512       &        27.6 $\pm$   2.8   	&	 0.5		   &  2.3          \\
     60032-01-12-000   &      52182.61618  & 52182.61667       &        13.6 $\pm$   1.5   	&	 0.9		   &  2.7          \\
      60032-01-14-01   &      52214.31827  & 52214.31882       &        18.9 $\pm$   1.4   	&	 1.3		   &  3.2          \\
     60032-01-18-00G   &      52273.69081  & 52273.69130       &        15.4 $\pm$   1.2   	&	 0.9		   &  2.0          \\
     60032-01-20-000   &      52283.01851  & 52283.01896       &        23.0 $\pm$   2.0   	&	 1.3		   &  2.3          \\
      60032-01-20-01   &      52283.53362  & 52283.53417       &        27.7 $\pm$   1.2   	&	 1.5		   &  2.4          \\
      60032-05-01-00   &      52286.05404  & 52286.05451       &         2.6 $\pm$   2.4   	&	 0.6		   &  1.9          \\
      60032-05-02-00   &      52286.55466  & 52286.55519       &        17.6 $\pm$   1.7   	&	 0.9		   &  2.0          \\
      60032-05-04-00   &      52287.52190  & 52287.52233       &         9.2 $\pm$   1.7   	&	 1.0		   &  2.0          \\
      60032-05-06-00   &      52288.51431  & 52288.51476       &        27.0 $\pm$   1.5   	&	 1.4		   &  2.1          \\
      60032-05-07-00   &      52288.97438  & 52288.97489       &         6.2 $\pm$   2.0   	&	 0.8		   &  1.9          \\
      60032-05-07-01   &      52289.29282  & 52289.29320       &        15.4 $\pm$   2.6   	&	 0.9		   &  1.9          \\
      60032-05-09-00   &      52289.97694  & 52289.97737       &        17.2 $\pm$   2.8   	&	 0.9		   &  2.1          \\
      60032-05-18-00  &       52390.21340 &  52390.21392       &         23.9 $\pm    1.2 $	&	 2.5		   &  3.0          \\
     60032-05-23-000   &      52646.77066  & 52646.77097       &         14.0 $\pm$   0.8  	&	 0.9		   &  2.4          \\
      91024-01-30-10  &       53688.95191 &  53688.95234       &         14.9 $\pm    1.8 $	&	 0.9		   &  4.3          \\
      91152-05-02-00   &      53919.07399  & 53919.07437       &        18.8 $\pm$   1.5   	&	 1.3		   &  4.0          \\
      92023-01-29-10  &       54050.90204 &  54050.90238       &         22.8 $\pm    3.5 $	&	 0.5		   &  2.9          \\
      92023-01-31-10  &       54054.24902 &  54054.24948       &         20.5 $\pm    1.5 $	&	 1.1		   &  2.9          \\
      70036-01-02-010 &       54271.04381 &  54271.04432       &         13.8 $\pm    0.8 $	&	 1.1		   &  3.1          \\
      70036-01-02-00   &      54272.09180  & 54272.09229       &        27.6 $\pm$   2.9   	&	 0.6		   &  3.2          \\
      93091-01-01-000 &       54371.71897 &  54371.71937       &         28.4 $\pm    1.8 $	&	 1.3		   &  2.0          \\
      93087-01-24-10  &       54522.68638 &  54522.68680       &         21.7 $\pm    1.7 $	&	 1.1		   &  2.5          \\
      93091-01-02-00   &      54523.57841  & 54523.57893       &         8.9 $\pm$   2.2   	&	 0.8		   &  2.8          \\
      93087-01-04-20  &       54678.26783 &  54678.26838       &         22.6 $\pm    2.0 $	&	 1.4		   &  2.5          \\
       94310-01-01-00   &     54904.83290  &   54904.83362      &        22.5 $\pm$  1.6   	&	 1.1		   &  2.6          \\
      94310-01-03-000 &       55079.21966 &  55079.22008       &         31.0 $\pm    1.7 $	&	 1.0		   &  2.4          \\

\hline          
\end{tabular}   
\medskip        
\\                  
\label{convexity}     
\end{table*}

\begin{table*}
\small
\caption{Properties of X-ray bursts.}
\centering
\begin{tabular}{ l  c  l  c }
\hline
\multicolumn{4}{l}{\bf 1. Results from simulations:} \\
\hline
{\bf\em a.} Low-latitude ignition & $\implies$ & $\mathcal{C} > 0$ & (1,2) \\
\phantom{\bf\em a.} High-latitude ignition & $\implies$ & $\mathcal{C} > 0$ {or $\mathcal{C} < 0$} & \\
&...&&\\
{\bf\em b.} Low-latitude ignition & $\implies$ & Short rising time & (1,2) \\
\phantom{\bf\em b.} High-latitude ignition & $\implies$ & Long/Short rising time & \\
\hline
\multicolumn{4}{l}{\bf 2. Results from observations:}\\
\hline
{\bf\em a.} mHz QPOs & $\implies$ & $\mathcal{C} > 0$ & (3) \\
\phantom{{\bf\em a.}} no mHz QPOs & $\implies$ & $\mathcal{C} > 0$ {or $\mathcal{C} < 0$} & \\
&...&&\\
{\bf\em b.} mHz QPOs & $\implies$ & Short rising time & (3) \\
\phantom{\bf\em b.} no mHz QPOs & $\implies$ & Long/Short rising time &  \\
\hline
\multicolumn{4}{l}{\bf 3. The statements {\em a} or {\em b} are logically equivalent to:}\\
\hline
\phantom{{\bf\em a.}} mHz QPOs & $\implies$ & Low-latitude ignition & \\
\phantom{{\bf\em b.}} no mHz QPOs & $\implies$ & Low-/High-latitude ignition & \\
\hline

\multicolumn{4}{l}{\footnotesize{References: (1) \citet{Maurer08}; (2) \citet{mah15}; (3) This paper.}}

\end{tabular}
\end{table*}

\medskip

\section{Discussion}

Using data from the full RXTE archive we found that all type I X-ray bursts associated with the mHz QPOs (39 in total) in 4U 1636--53 have positive convexity. We did not find a single case in our sample of an X-ray burst with negative convexity associated to a mHz QPO. The probability that this happens only by chance is less than 6$\times$10$^{-4}$, corresponding to a significance level of $\sim$ 3.5 $\sigma$.

Using numerical simulations of the propagation of a burning front on the neutron-star surface, \citet{Maurer08} found that bursts that ignite at the equator always have positive convexity, whereas bursts that ignite at high latitude have both positive and negative convexity. \citet{mah15} confirmed this result in their simulations, and also found that the rising time of bursts that ignite at the equator is short, whereas the rising time is both short or long for bursts that ignite at high latitudes \citep[see also][]{Maurer08}. In Table 2 we summarise the results of \citet{Maurer08} and \citet{mah15}, statements {\bf\em 1a} and {\bf\em 1b}, together with our own findings, statements {\bf\em 2a} and {\bf\em 2b}. The last row in that Table shows the statements, {\bf\em 3}, that follow logically from either the {\em a} or the {\em b} statements.

Bursts with short rising time are likely fuelled by Helium \citep{fujimoto81}. The apparent connection between mHz QPO and bursts with short rising time (Figure \ref{rising}) suggests the possibility that mHz QPOs are due to marginally-stable nuclear burning of Helium on the neutron-star surface \citep{heger07}. However, there are as many bursts with a short rising time without mHz QPO as with mHz QPO (lower right corner of Figure \ref{rising}), which indicates that marginally-stable Helium burning can not be the only reason for the presence of mHz QPOs.

From Table 2 we can conclude that all the 39 bursts with positive convexity and short rising time that are associated with mHz QPOs ignited at the neutron-star equator. For, if bursts associated with mHz QPOs ignited anywhere on the neutron-star surface (therefore these 39 bursts would correspond to cases of positive convexity and either low- or high-latitude ignition in the analysis of \citet{Maurer08}), we would have expected to see also cases of mHz QPOs associated with bursts with negative convexity in our sample. While in this scenario bursts with positive convexity but no associated mHz QPO would have in principle ignited at high latitudes, some of them may also have ignited at the equator if, for instance, those bursts happened at an accretion rate in which marginally stable nuclear burning would not be at work \citep[e.g.,][]{heger07}. Also, in some cases a QPO might be present just before an X-ray burst, but we are unable to detect it either because we do not have enough data before a burst (e.g., if the data segment before the burst was too short), or because the data are not of sufficient quality to detect the QPO significantly (e.g., if some PCU detectors were not operating during that observation).

The simplest scenario that follows from this is that the marginally-stable burning (that produces the QPO) and the unstable burning (that produces the burst) take place at the same physical location. There should still be enough fuel at the equator to trigger a burst after the mHz QPOs if, similar to the case of unstable burning at high luminosity \citep[e.g.,][]{para88,muno00,corn03,heger07}, marginally-stable burning consumes only a fraction of the fuel on the surface of the neutron star. We cannot discard, however, more complex scenarios in which the sites of marginally-stable and unstable burning are physically disconnected, mHz QPOs and bursts with positive convexity happen at any latitude, but some other mechanism ensure that mHz QPOs and bursts with positive convexity are causally connected.

\citet{fujimoto81} proposed that the thermal stability and burst ignition of a neutron star actually depends on the accretion rate per unit area, $\dot{m}$, instead of the global accretion rate. The quantity $\dot{m}$ needs not to be the same everywhere on the neutron-star surface \citep[e.g.,][]{fujimoto81,review}. During accretion, the infalling matter first reaches the equator and then spreads over the whole surface of the neutron star, therefore $\dot m$ will be higher at the equator than at high latitudes. If the mHz QPOs happen at the equator, the local accretion rate per unit area, $\dot{m}$, would also be the key parameter that determines whether marginally stable nuclear burning on the neutron-star surface takes place: when nuclear burning occurs around the equator, $\dot{m}$ is high enough to trigger the mHz QPOs, while there are no mHz QPOs when the nuclear burning happens at high latitudes where $\dot{m}$ is below the threshold value to trigger the marginally-stable nuclear burning process. The fact that the distribution of the persistent flux of observations with and without mHz QPOs is consistent with being the same further enhances the argument that it is the local accretion rate $\dot{m}$ that triggers mHz QPOs. This picture is similar to the one proposed in \citet{heger07} in which the accreted fuel that is responsible for the marginally-stable nuclear burning is confined at a certain burning depth, where the local accretion rate could be much higher than the global accretion rate. This scenario is able to bridge the gap between the high accretion rate required for triggering the mHz QPOs in the models and the relatively low accretion rate implied from observations. 

\section*{Acknowledgments}

This research has made use of data obtained from the High Energy Astrophysics Science Archive Research Center (HEASARC), provided by NASA's Goddard Space Flight Center. This research made use of NASA's Astrophysics Data System. LM is supported by China Scholarship Council (CSC), grant number 201208440011. DA acknowledges support from the Royal Society.

\clearpage


\label{lastpage}

\end{document}